\documentclass[reprint,showpacs,preprintnumbers,amsmath,amssymb,prc,floatfix]{revtex4-1}
\usepackage{color}
\usepackage{adjustbox}
\usepackage[mathcal]{eucal}
\usepackage{graphicx}
\usepackage{dcolumn}
\usepackage{bm}
\usepackage{xcolor}
\RequirePackage[colorlinks,citecolor=red,linkcolor=blue,anchorcolor=blue,filecolor=blue,urlcolor=blue]{hyperref}

\begin{document}
\title{Implementation of the microscopic nuclear potential in the coupled channels calculations to study the fusion dynamics of Oxygen based reactions}
\author{N. Jain$^{1}$}
\email{nishujain1003@gmail.com}
\author{M. Bhuyan$^{2}$}
\email{bunuphy@um.edu.my}
\author{Raj Kumar$^1$}
\email{rajkumar@thapar.edu}

\bigskip
\affiliation{$^1$Department of Physics and Materials Science, Thapar Institute of Engineering and Technology, Patiala 147004, India}
\affiliation{$^2$Center for Theoretical and Computational Physics, Department of Physics, Faculty of Science, Universiti Malaya, Kuala Lumpur 50603, Malaysia}
\date{\today}

\begin{abstract}
\noindent
In the present work, we have incorporated the microscopic relativistic nuclear potential obtained from recently developed relativistic R3Y NN potential in the coupled channels code CCFULL to study the fusion dynamics. The R3Y NN-potential and the densities of interacting nuclei are obtained for the relativistic mean-field approach for the NL3$^*$ parameter set. It is to be noted that the R3Y NN potential can be expressed in terms of masses of the mesons and their couplings by considering the meson degrees of freedom within the relativistic mean field, which has a form similar to the widely used M3Y potential. We focused on the fusion cross-sections for $Oxygen$-based reactions with targets from different mass regions of the periodic table i.e. $^{16}$O + $^{24}$Mg, $^{18}$O + $^{24}$Mg, $^{16}$O + $^{148}${Sm}, $^{16}$O + $^{176}${Hf}, $^{16}$O + $^{176}$Yb, $^{16}$O + $^{182}${W}, and $^{16}$O + $^{186}$W. A comparison is also made with the ones calculated using the nuclear potential obtained from the traditional Woods-Saxon potential and the widely used M3Y NN potential within CCFULL. The coupled channel calculations are performed with shape and rotational degrees of freedom to examine the fusion enhancement at below-barrier energies. It is observed from the calculations that the fusion cross-sections obtained using R3Y NN potential with rotational degrees of freedom are found to be more consistent with the experimental data than those for the M3Y and Woods-Saxon potentials mainly at below barrier energies.
\end{abstract}  
\maketitle
\section{Introduction} \label{introduction}
\noindent
The fusion reactions of heavy nuclei have been thoroughly studied to comprehend the quantum tunnelling phenomenon in the detailed study of many-body systems \cite{dasg98,bala98,cant06,back14}. Analyzing data from fusion reactions gives ample insight into synthesizing new elements or superheavy nuclei to extend the periodic table.  The 1-D Barrier Penetration Model (BPM) is used to compute the fusion cross-section in the most basic description of nuclear fusion reactions \cite{bohr39}. The single fusion barrier can be viewed as breaking into the distribution of barriers. However, the coupling effects resulting from the vibrational, rotational, and neutron transfer degrees of freedom \cite{bala98,esbe87,hagi12,gaut19,rajb16,rajb14,jish22} must be considered when the collision energy is in the sub-barrier energy regime. The description of the fusion cross-sections has been done with the help of several coupled channels (CC) codes \cite{dasso87,fern89,hagi99,newt01} out of which the CCFULL code is frequently employed in fusion reaction cross-section calculations \cite{hagi97,lin09,mei21}. Furthermore, the study of fusion characteristics at extreme sub-barrier energies is crucial for understanding the reaction mechanisms in astrophysics and synthesizing superheavy nuclei \cite{rowl91,vand92,bala98}. While nuclear structure effects dominate below the Coulomb barrier, the centrifugal potential suppresses these effects near or above the barrier. Nevertheless, the nuclear interaction component still needs to be understood \cite{raj11,jain12,raj2011,jain14}. The study of fusion dynamics is crucially dependent on understanding the pivotal role played by nuclear potential. Interestingly, the nuclear potential determines both the shape of potential and the height of the Coulomb barrier as prescribed by the well-known Wong formula \cite{wong73}. Furthermore, the nuclear potential of the ground state affects the nuclear coupling to the excitation states of the colliding nuclei. Consequently, several nuclear potentials have been used in the coupled-channel approach to explain the fusion of heavy ions using 1-D BPM \cite{krap79,bloc77,bass74}. \\ \\
Various theoretical approaches have been developed to obtain and explain the nuclear potential between two colliding nuclei. One well-known method is the double-folding approach, where the ion-ion optical potential is derived by integrating an effective NN interaction and nuclear densities. This approach has been successfully applied to study nuclear clusters, proton-radioactivity, fusion, elastic and inelastic scattering \cite{satc79,bert77,khoa16,long08,bhuy18,bhuy20}. The nuclear densities, obtained using the Woods-Saxon ansatz, two-parameter Fermi (2pF), three-parameter Fermi (3pF), Skyrme Hartree Fock (SHF) models, and the M3Y effective NN potential, are used to calculate the nuclear potential within the double-folding approach \cite{cham02,fold}, along with relevant references. The non-relativistic M3Y NN potential is expressed as the sum of one pion exchange potential (OPEP) and Yukawa terms, fitted to reproduce the G-matrix elements on an oscillator basis. In essence, the two crucial inputs in the double-folding approach are the densities of interacting nuclei and the NN-potential (M3Y), obtained from two independent approaches. On the other hand, the nuclear potential for the interacting nuclei is generated via the Woods-Saxon (WS) potential in the CCFULL code \cite{hagi99}, a simple empirical formula fitted for the known region of the nuclear chart. The Woods-Saxon potential, composed of depth, range, and diffuseness parameters, is widely used for approximating the shape of the nuclear component of nucleus-nucleus interactions \cite{gan21,gaut15}. The parameters of the WS potential are chosen to fit the experimental fusion cross-section, mainly at above-barrier energies. However, the synthesis of exotic nuclei far from the $\beta$-stable region of the nuclear chart requires the adoption of a microscopic nuclear potential to study the fusion dynamics of the exotic region of the nuclear chart, as discussed in our previous work \cite{rana22} and references therein.

At the microscopic level, the Skyrme-Hartree-Fock (SHF) and the relativistic mean-field (RMF) model \cite{Rein00}, can construct a nuclear interaction potential potential from a double folding procedure. At low energy, nucleon-nucleon interactions are instantaneous, allowing for the concept of an interaction potential via an intermediate particle. The resulting nucleon-nucleon (NN) interaction potential, derived through particle exchange, is a significant tool for the understanding of nuclear forces and structure properties \cite{wein90,kapl88}. A double-folding procedure is used to calculate an optical potential between two interacting nuclei using this fundamental NN interaction \cite{sing12,sahu14,bhuy14}. Recently, the relativistic R3Y effective NN potential has been used in terms of meson masses and coupling constants and derived from the self-consistent relativistic mean-field (RMF) approach for a particular parameter set. Further, the densities are also calculated for the same parameter sets for both interacting nuclei, which are simultaneously used in the double-folding model. This enables the advancement of the theoretical approach to obtain the nuclear potential through meson interaction and also keeps the consistency of parameterization for NN potential and densities. It is to be noted that RMF models have proven to be highly predictive in the structural features of finite nuclei for $\beta$-stable and also highly isospin asymmetry regions of the nuclear chart \cite{satc79,sing12,sahu14,lahi16,sero86,bhuy15,bhuy18,bhuy18a,bhuy20,ring96,rein89}.\\ \\
Furthermore, applying the relativistic NN-interaction potential along with nuclear densities from the RMF formalism has proven successful in describing various nuclear phenomena such as clustering, proton radioactivity, and nuclear fusion \cite{sing12,sahu14,bhuy14}. Combining the relativistic R3Y potential with the RMF density is a compelling approach for investigating low-energy fusion reactions across various systems. In our previous works, the M3Y and R3Y nucleon-nucleon potentials were used to explore fusion hindrance phenomena in selected Ni-based reactions and to calculate cross-sections for synthesizing heavy and superheavy nuclei. Specifically, the fusion and/or capture cross-sections calculated using the M3Y NN potential were compared to the relativistic R3Y NN potential for the NL3$^*$ parameter set in earlier studies. A similar approach was employed using the $\ell$-summed Wong formula in our previous work \cite{bhuy20,bhuy18,bhuy22} and references therein. These studies concluded that the relativistic R3Y potential provides better agreement with experimental data than the M3Y potential \cite{sing12,sahu14,bhuy18,bhuy20,shil21,josh22}. The novelty of the present work lies in obtaining the nucleus-nucleus potential from a double-folding model using a relativistic potential that has not been used as input for coupled channels calculations. Furthermore, incorporating proper structural effects within the microscopic approach using RMF cannot be underestimated or undervalued, as it plays a vital role in reaction studies. Therefore, the involvement of rotational degrees of freedom with the microscopic R3Y NN interaction potential is considered for the target nuclei, which is limited to spherical symmetry in our previous work \cite{bhuy18,bhuy20,shil21}. Hence, the present study represents a significant step forward in developing the nuclear potential obtained using the RMF formalism and its application within the coupled channel approach for studying fusion dynamics. The study includes reactions, i.e., $^{16}$O+$^{24}$Mg, $^{18}$O+$^{24}$Mg, $^{16}$O+$^{148}${Sm}, $^{16}$O+$^{176}${Hf}, $^{16}$O+$^{176}$Yb, $^{16}$O+$^{182}${W}, and $^{16}$O+$^{186}$W, covering different mass regions. The fusion cross-sections for these reactions are analyzed at sub-barrier energies using the M3Y and microscopic R3Y NN potentials for the NL3$^*$ parameter set, and further compared with the case of the static Woods-Saxon potential. A possible comparison will be made between the theoretical results and the experimental data \cite{tabo78,leig88,leig95,rajb16,trot05} for the considered reaction systems. Thus, implementing the microscopic nuclear potential in the CCFULL code to study fusion dynamics becomes crucial and exciting.\\ \\
This paper is organized as follows: Sec. \ref{theory} gives an overview of the theoretical formalism used for calculations. Sec. \ref{results} shows the findings of the coupled channels calculations. The results and discussions of the current work are summarised in Sec. \ref{result}.

\section{Theoretical Formalism} \label{theory} 
\noindent
The Coupled channels approach (CCFULL) is employed in the present study, which offers a reasonable understanding of the nuclear fusion dynamics at energy near the barrier. Instead of focusing on single barrier penetration, this approach considers multidimensional barrier penetration. This method considered the influence of coupling between relative motion and intrinsic degrees of freedom of the interacting nuclei, mainly for calculating mean angular momenta and the fusion cross-sections of the compound nucleus \cite{bala98,hagi12,hagi99}. The standard approach for addressing the impacts of the coupling between relative motion and intrinsic degrees of freedom on fusion is to numerically solve the coupled channels equations, which include all relevant channels \cite{Esbe87a,rumi99}. The coupled channels equations solved numerically within the CCFULL are given as:
\begin{eqnarray}
\Bigg[\frac{-\hbar^2}{2\mu} \frac{d^2 }{dr^2} &+& \frac{J(J+1)\hbar^2}{2\mu r^2}+\frac{Z_P Z_T e^2}{r}+V_{N}+\epsilon_n -E_{c.m.}\Bigg] \nonumber \\
&& \times \psi_n (r) + \sum_m V_{nm} (r)\psi_m (r) =0. 
\label{cc1}   
\end{eqnarray}
Here $r$ represents the radial part of relative motion between the colliding nuclei, and $\mu$ is known as the reduced mass of the interacting system. $\epsilon_n$ is the excitation energy for the $n^{th}$ channel and $E_{c.m.}$ is the bombarding energy in the center of the mass frame. $V_{N}$ is the nuclear potential, and $V_{nm}$ represents the matrix elements of the coupled Hamiltonian. Since there are several Coupled channel equations, their dimension is also significant. To reduce the dimension of coupled channels equations, rotating frame approximation or no-Coriolis approximation is used \cite{hagi99,hagi98,hagi12}. The CC equations with non-linear coupling are significant in studying the heavy-ion fusion reactions mainly at sub-barrier energies. All these sets of non-linear coupling are taken into account.

The boundary condition for the incoming wave \cite{land84} is also necessary for the solution of the Coupled channels equation as it is sensitive to the potential pocket of the interaction fusion barrier. The incoming wave of the entrance channel is at the minimum position (r = $r_{min}$) of the barrier and the outgoing wave of other channels is at an infinite position. By involving the effect of the prominent intrinsic channels, the fusion cross-sections are calculated as given below:
\begin{eqnarray}
\sigma_J (E)=\sigma_{fus} (E)=\frac{\pi}{k_{0}^{2}}\sum_{J}(2J+1)P_J (E).
\label{fusion}
\end{eqnarray}
Here, the iso-centrifugal approximation is used, and the total angular momentum {\it J} is substituted in place of $\ell$ for each channel by using the following equation:
\begin{eqnarray}
   \langle{\ell}\rangle&=&\sum_{J} J \sigma_J (E) / \sum_{J} \sigma_J (E)
    =\Bigg(\frac{\pi}{k_{0}^{2}}\sum_{J} J(2J+1)P_J (E)\Bigg) \nonumber \\
&&   \bigg / \Bigg(\frac{\pi}{k_{0}^{2}}\sum_{J} (2J+1)P_J (E)\Bigg),   
 \label{am}
\end{eqnarray}
where $P_J (E)$ is the total transmission coefficient. 
The crucial ingredient of the coupled channels approach is the nucleus-nucleus interaction potential, which is taken as a Woods-Saxon form for the known region of the nuclear chart.
\begin{equation}
    V_{N}= \frac{-V_0}{1+\exp\Big[\big(r_0-R_0\big)/a_0\Big]},  
    \\ \label{ws}
\end{equation}
where $V_0$, $r_0$, and $a_0$ are the nuclear potential parameters. In heavy-ion fusion reactions, the parameter of nuclear potential is at equal footing as that of nuclear structure degrees of freedom. Parallel to the traditional Woods-Saxon potential, here we employ the nuclear potential obtained from widely used M3Y \cite{satc79} and relativistic R3Y \cite{sing12,sahu14,bhuy14} NN potential for the NL3$^*$ parameter set. It is worth mentioning that the nuclear potential obtained from relativistic mean-field is first time introduced in the CCFULL for the study of fusion dynamics. 

The recently developed R3Y NN potential can be obtained from the static solution of the field equations for mesons \cite{sing12,bhuy14,bhuy18}, and can be written as:
\begin{eqnarray}
V_{\mbox{eff}}^{R3Y}(r)&=&\frac{g^2_\omega}{4\pi}\frac{e^{-m_\omega r}}{r}+\frac{g^2_\rho}{4\pi}\frac{e^{-m_\rho r}}{r} -\frac{g^2_\sigma}{4\pi}\frac{e^{-m_\sigma r}}{r} \nonumber \\
&+&\frac{g^2_2}{4\pi}re^{-2m_\sigma r} +\frac{g^2_3}{4\pi}\frac{e^{-3m_\sigma r}}{r}+J_{00}(E)\delta(r),
\label{r3y}
\end{eqnarray}
Here, the parameters $g_\sigma$, $g_\omega$, $g_\rho$ denote the respective coupling constants of the mesons having masses $m_\sigma$, $m_\omega$ and  $m_\rho$, respectively. The $g_2$ and $g_3$ are the coupling constants of the non-linear terms of the self-interacting $\sigma$ field. The $J_{00}$ is the one-pion exchange potential (OPEP) and details can be found in Ref. \cite{satc79}. Here we have used a revisited version of the widely used NL3 force \cite{lala97}, so-called NL3$^*$ parameter set \cite{lala09}. In the same pattern, the M3Y NN potential can be expressed as, 
\begin{equation}
    V_{\mbox{eff}}^{M3Y}(r)=7999\frac{e^{-4r}}{4r}-2134\frac{e^{-2.5r}}{2.5r}+J_{00}(E)\delta(r). 
    \label{m3y}
\end{equation}
Here, the range unit is in fm and the strength is in MeV. More details of Eqs. (\ref{r3y}) and (\ref{m3y}) can be found in Refs. \cite{bhuy18,sing12,sahu14,schi51,satc79}. The interaction potential between the projectile and target nuclei, considering their respective calculated nuclear densities $\rho_p$ and $\rho_t$ with the RMF approach for NL3$^*$ parameter, can be determined using 
\begin{eqnarray}
V_{n}(\vec{R}) = \int\rho_{p}(\vec{r}_p)\rho_{t}(\vec{r}_t)V_{eff}
\left( |\vec{r}_p-\vec{r}_t +\vec{R}| {\equiv}r \right) 
d^{3}r_pd^{3}r_t,
\label{DF}
\end{eqnarray}
the double-folding procedure \cite{satc79} with the M3Y and relativistic R3Y interaction potentials proposed in Refs. \cite{bhuy18,sing12,sahu14}. Additionally, single nucleon exchange effects can be included through a zero-range pseudo-potential.

The total nuclear interaction potential between the projectile and target nuclei can be obtained by combining the Coulomb potential with the nuclear interaction potential $V_n(R)$ obtained from Eq. (\ref{DF}), which is the main ingredient in the Coupled channels approach (CCFULL). One can generate the nuclear coupling Hamiltonian by changing the target radius in the nuclear potential \cite{dasg98} to a dynamical operator,
\begin{equation}
   R_0 \rightarrow R_0 + \hat{O}= R_0 + \beta_2R_TY_{20}+\beta_4R_TY_{40},
\end{equation}
where $R_T$ is $r_{coup}A^{1/3}$ and $\beta_2$ and $\beta_4$ are the quadrupole and hexadecapole deformation parameters of the deformed target nucleus, respectively. Thus, the nuclear coupling Hamiltonian is given by
\begin{equation}
    V_{N}(r,\hat{O})= \frac{-V_0}{1+\exp\Big[\big(r_0-R_0-\hat{O}\big)/a_0\Big]}. 
    \label{ws1}
\end{equation}

In order to connect the $|n\rangle=|I0\rangle$ and $|m\rangle=|I'0\rangle$ states of the target's ground rotational band, we need matrix elements of the coupling Hamiltonian. These are readily accessible using matrix algebra \cite{Kerm93}. In this algebra, the eigenvalues and eigenvectors of the operator $\hat{O}$, which satisfies
\begin{equation}
    \hat{O}\mid \alpha > = \lambda_\alpha\mid \alpha > 
\end{equation}
This is implemented in the CCFULL program by diagonalizing the matrix $\hat{O}$, whose elements are given by 
\begin{eqnarray}
\begin{aligned}
    \hat{O}_{II'}= \sqrt{\frac{5(2I+1)(2I'+1)}{4\pi}}\beta_2R_T
\begin{pmatrix}
I & 2 & I'\\
0 & 0 & 0 
\end{pmatrix}^2 \\
+ \sqrt{\frac{9(2I+1)(2I'+1)}{4\pi}}\beta_4R_T
\begin{pmatrix}
I & 4 & I'\\
0 & 0 & 0 
\end{pmatrix}^2.  
\end{aligned}
\end{eqnarray}
The nuclear coupling matrix elements are then evaluated as
\begin{eqnarray}
V_{nm}^{(N)}&=& <I0\mid V_N(r,\hat{O})\mid I^{'}0> - V_{N}^{(0)}(r)\delta_{n,m}, \nonumber \\
&&  = \sum_\alpha <I0\mid \alpha><\alpha\mid I^{'}0> V_N(r,\lambda_\alpha)\nonumber \\
&& -V_{N}^{(0)}(r)\delta_{n,m}.   
\label{am1}
\end{eqnarray}
The last term is included in the equation to avoid the diagonal component from being counted twice. The CCFULL program incorporates the Coulomb interaction of the deformed target up to the second order in $\beta_2$ and the first order in $\beta_4$. Although the higher-order couplings of the Coulomb interaction have a relatively minor impact, unlike the nuclear couplings. The matrix elements of the Coulomb interaction potential can be expressed as follows:
\begin{eqnarray}
    \begin{split}
        V^C_{R{(I,I')}}=\frac{3Z_PZ_TR^2_T}{5r^3}\sqrt{\frac{5(2I+1)(2I'+1)}{4\pi}}\times \\
        \Bigg( \beta_{2}+\frac{2}{7}\beta^2_2\sqrt{\frac{5}{\pi}}\Bigg)\begin{pmatrix}
I & 2 & I'\\
0 & 0 & 0 
\end{pmatrix}^2 \\+\frac{3Z_PZ_TR^4_T}{9r^5}\sqrt{\frac{9(2I+1)(2I'+1)}{4\pi}}\times\\
\Bigg( \beta_{4}+\frac{9}{7}\beta^2_2\Bigg)\begin{pmatrix}
I & 4 & I'\\
0 & 0 & 0 
\end{pmatrix},
    \end{split}
\end{eqnarray}
for rotational coupling. These coupled channel equations are used to calculate the fusion cross-section of the compound nucleus by assuming the rotational degrees of freedom as discussed in Sec. \ref{results}.

\section{Results and Discussions}
\label{results} \noindent
The coupled channels calculations adequately account for several degrees of freedom such as collective surface vibrations, rotations, and neutron transfer to reasonably explain the fusion cross-section with lower excitation energy. In the present study, the Coupled channel calculations with microscopic external potential are performed by using the following steps:
\begin{enumerate}
     \item The nucleus-nucleus potential is obtained from a double folding model using the relativistic potential that has not been used as input to coupled channels calculations, including low-lying rotational states of the deformed target nucleus among the channels.
    \item The structural bulk properties such as binding energy, nuclear radii, density distribution, quadrupole ($\beta_2$) and hexadecapole ($\beta_4$) quadrupole are obtained from relativistic mean field formalism, which gives a picture of the nuclei that involve in the reaction. 
    \item The density, and the shape's degrees of freedom i.e., quadrupole ($\beta_2$) and hexadecapole ($\beta_4$) are incorporated through nuclear potential to calculate the fusion cross-section within the coupled channels calculations. 
    \item The rotational degrees of freedom up to $4^+$ states are considered to examine the fusion enhancement at below-barrier energies.
\end{enumerate}
It is worth mentioning that the primary reason for choosing the external nuclear potential obtained from RMF is that it provides valuable information on bulk nuclear properties such as binding energy, charge distributions, and single-particle energy levels. Further details on the applicability of the RMF model with various parameterizations can be found in Refs. \cite{sing12,bhuy14,bhuy18} and references therein. In parallel, the Woods-Saxon (WS) potential within CC calculations is also used for comparison. The reactions, namely, $^{16}$O + $^{24}$Mg,  $^{18}$O + $^{24}$Mg, $^{16}$O + $^{148}${Sm}, $^{16}$O + $^{176}${Hf}, $^{16}$O + $^{176}$Yb,  $^{16}$O + $^{182}${W}, and $^{16}$O + $^{186}$W are considered within the CCFULL code to estimate the fusion cross-sections. Rotational degrees of freedom have been considered for the target nuclei \cite{tabo78,leig88,leig95,rajb16,trot05}, while $^{16}$O and $^{18}$O are treated as spherical in the present analysis.\\

\begin{table} 
\caption{\label{table1} The Woods-Saxon (WS) parameters ($V_0$, $r_0$, $\&$ $a_0$), quadrupole ($\beta_2$) and hexadecapole ($\beta_4$) deformation calculated using RMF  (with NL3$^*$ parameter set) are listed in table given below. The excitation energy corresponds to quadrupole deformation of the target nuclei is taken from Ref. \cite{rama01}.}
\renewcommand{\tabcolsep}{0.1cm}
\renewcommand{\arraystretch}{1.6}
\begin{tabular}{|c|ccc|ccc|ccc|ccc|}
\hline \hline 
System & \multicolumn{3}{c|}{Woods-Saxon Potential} 
 & \multicolumn{3}{c|}{Deformation Target} \\
& $V_0$ & $r_0$ & $a_0$ & $E_2^+$ & $\beta_2$ & $\beta_4$ \\
& (MeV) & (fm) & (fm) & (MeV) & & \\
\hline
$^{16}$O+$^{24}$Mg  & 45.93& 1.17 & 0.61 & 1.368 & 0.416& 0.004 \\
$^{18}$O+$^{24}$Mg  & 47.20& 1.17& 0.62 & 1.368 & 0.416& 0.004\\  
$^{16}$O+$^{148}$Sm & 62.20& 1.17& 0.65 & 0.550  & 0.112& 0.049\\
$^{16}$O+$^{176}$Hf & 63.63& 1.18 & 0.65 & 0.088 & 0.284& -0.041 \\
$^{16}$O+$^{176}$Yb & 60.00  & 1.17& 0.65 & 0.082 & 0.298& -0.057 \\
$^{16}$O+$^{182}$W  & 63.99& 1.17& 0.65 & 0.100 & 0.273& -0.056 \\
$^{16}$O+$^{186}$W  & 70.00  & 1.18 & 0.65 & 0.122 & 0.241& -0.094\\
\hline \hline 
\end{tabular}
\end{table}
\begin{figure*}
\begin{center}
\includegraphics[width=160mm,height=60mm,scale=1.5]{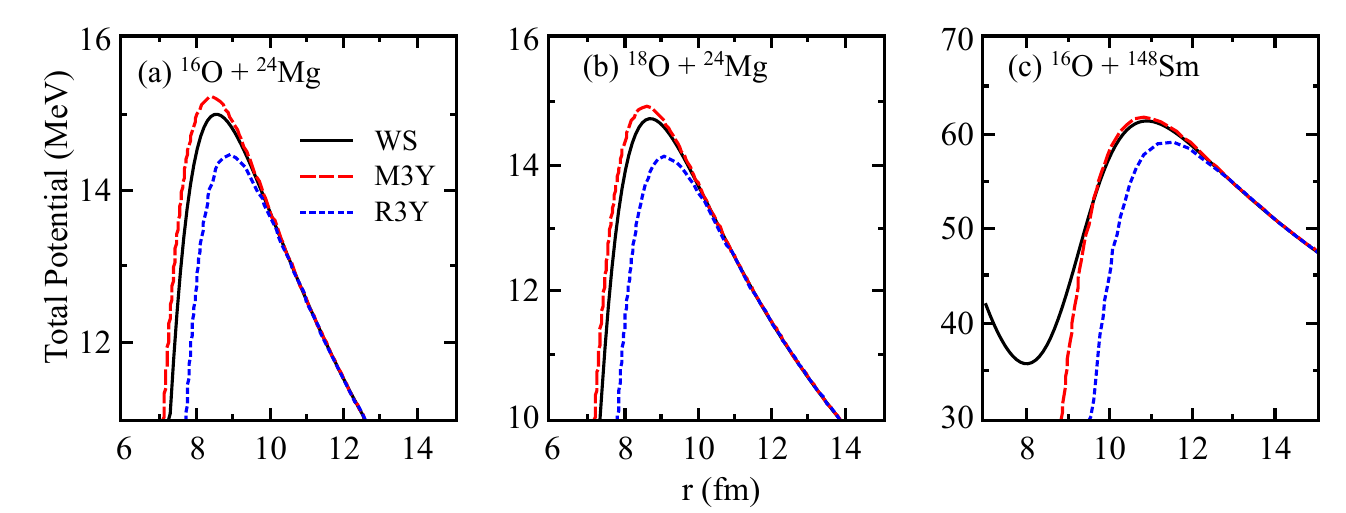}
\vspace{-0.3cm}
\caption{\label{fig1} (Color online) The total interaction potential as a function of radial separation $r$ for $^{16}$O+$^{24}$Mg,  $^{18}$O+$^{24}$Mg,  and $^{16}$O+$^{148}${Sm} reactions calculated using the M3Y (dashed red lines), R3Y (dotted blue lines) NN potential. The microscopic nuclear potential is further compared with the traditional WS (solid black line) potential. See the text for details.}
\end{center}
\end{figure*}
\begin{figure*}
\begin{center}
\includegraphics[width=175mm,height=77mm,scale=1.5]{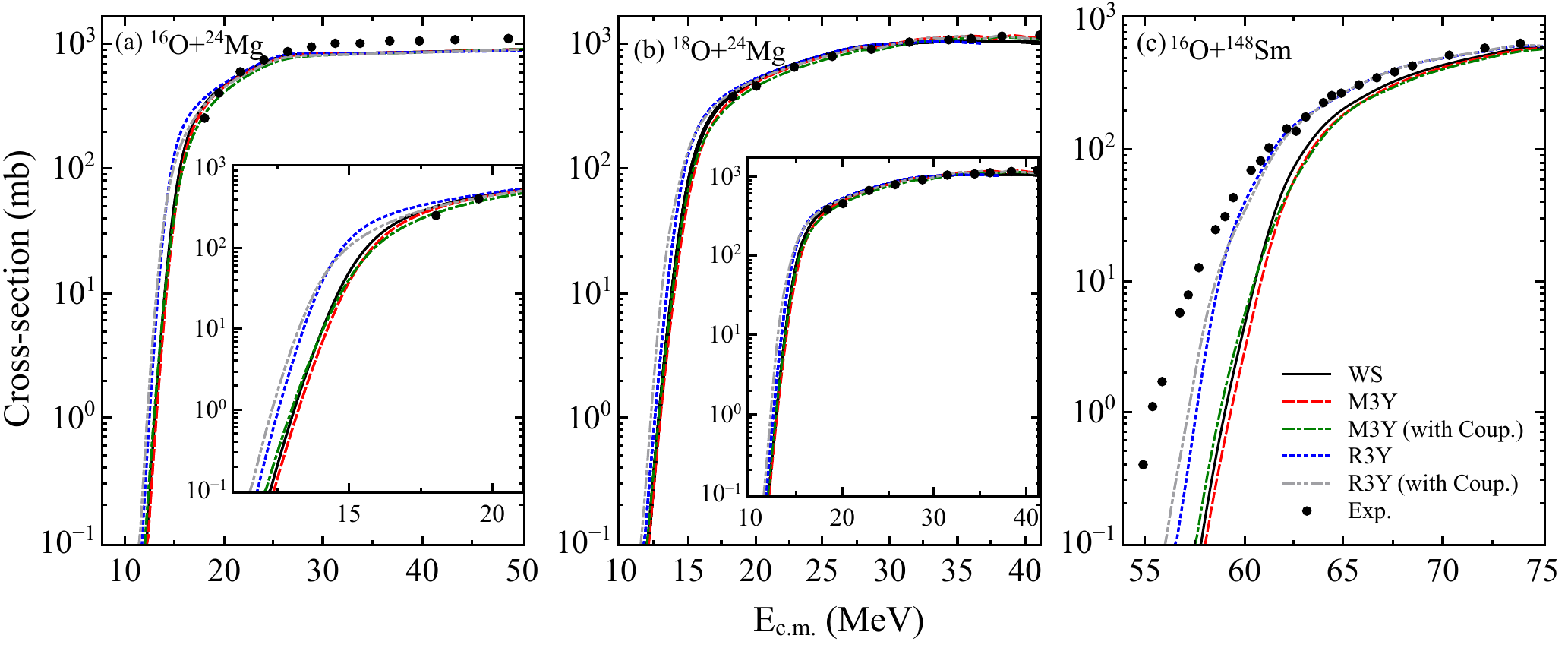}
\vspace{-0.3cm}
\caption{\label{fig2} (Color online){ The fusion cross-sections as a function of $E_{c.m.}$ (MeV) for Woods-Saxon potential (solid black line), R3Y (dotted blue line) and M3Y (dashed red line) NN interactions with NL3$^*$ parameter set using Coupled channels code. The corresponding cross-section for the 4$^+$ excited state for R3Y (double-dotted dashed grey line) and  M3Y(dashed-dotted red line) potential along with the inclusion of $\beta_2$, and $\beta_4$ deformation. The calculated results are compared with the experimental data \cite{tabo78,leig95} for $^{16}$O+$^{24}$Mg, $^{18}$O+$^{24}$Mg, and $^{16}$O+$^{148}${Sm} reactions. See the text for details.}}
\end{center}
\end{figure*}
The Woods-Saxon (WS) parameterizations of Aky$\ddot{u}$zWinther potential (AW), the excitation energy corresponding to the first excitation state \cite{rama01} and the values of the deformation parameters (calculated using RMF with NL3$^*$ parameter set) are listed in Table \ref{table1}. The potential parameters of WS potential are chosen in such a way as to fit the experimental data at the above barrier energies for 1-D BPM. The fusion barrier characteristics i.e., barrier height and position, are obtained using the above-mentioned potentials, namely, WS, M3Y, and R3Y NN potential. The variation of total interaction potential, which is given by the sum of Coulomb and nuclear potential, at $\ell$ = 0$\hbar$ with the separation distance `$r$' is shown in the Fig. \ref{fig1} and inset of Fig. \ref{fig3} for the considered choice of reactions. The solid black and dotted blue line represents the interaction potential corresponding to WS and R3Y NN potential while the dashed red line represents the M3Y NN potential. It can be observed from these figures that the barrier formed in the case of M3Y NN potential and WS potential is relatively higher in comparison to the relativistic R3Y NN potential (as discussed in our previous work \cite{rana22}) for all the considered reactions. The results obtained from the above analysis suggest that the R3Y effective NN potential, formulated in terms of meson masses and coupling constants, generates a more attractive interaction potential than the M3Y NN and WS potential. The fusion cross-section of reaction systems is significantly influenced by the properties of the fusion barrier, highlighting the direct impact of barrier characteristics on the overall reaction dynamics. The higher the barrier height, the lower will be the fusion cross-section. The effect of the above-discussed potentials on the fusion cross-section is studied further with the help of coupled channel calculations.
\begin{figure*}
\begin{center}
\includegraphics[width=150mm,height=160mm,scale=1.5]{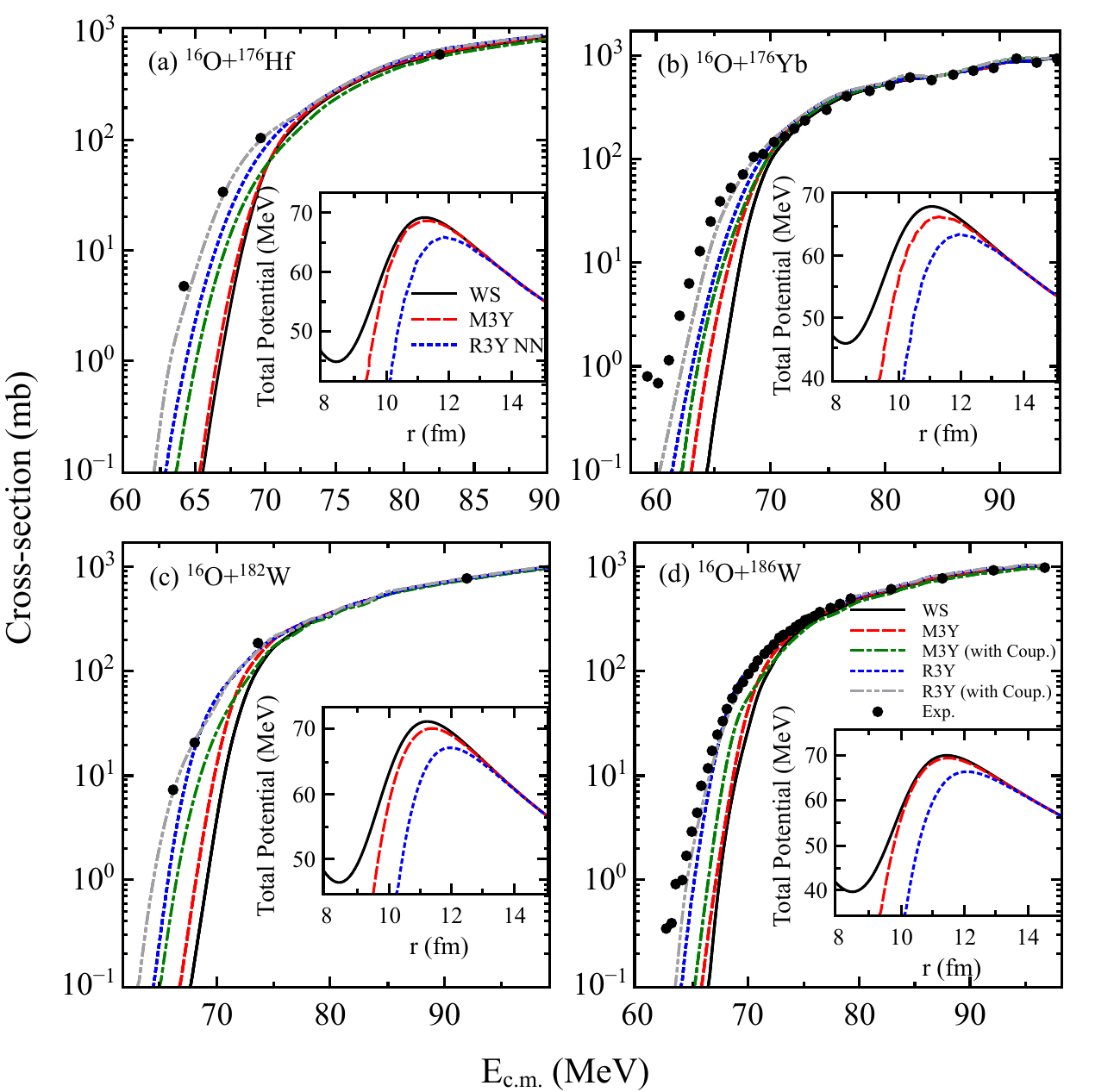}
\vspace{-0.3cm}
\caption{\label{fig3} (Color online) Same as for Fig. \ref{fig2}, but for $\beta_4 < 0$, namely, reaction $^{16}$O+$^{176}${Hf}, $^{16}$O+$^{176}$Yb,  $^{16}$O+$^{182}${W}, $^{16}$O+$^{186}$W. The experimental data are taken from Refs. \cite{rajb16,leig88,trot05}. See the text for details.}
\end{center}
\end{figure*}

Following Ref. \cite{kuma09}, we obtain the $\ell_{max}$ values from the sharp cut-off model \cite{beck81} by using experimental data at above barrier energies, wherever available and extrapolated for below barrier energies. As we mentioned above, the main aim of our work is to use the microscopic nuclear potential in the CCFULL code obtained from the recently developed R3Y NN potential. The calculated results are further compared with the cross-section obtained using traditional WS and M3Y NN potential. The RMF with NL3$^*$ parameter set is used to obtain the relativistic R3Y NN-potential using the density of the projectiles and targets, which are the main ingredients for nuclear potential. It is worth mentioning that the self-consistent relativistic mean-field formalism is successfully applied in the fusion hindrance reaction phenomena in our recent works \cite{bhuy20,bhuy18,shil21}. Furthermore, the RMF with NL3$^*$ force parameter could provide the relativistic flavour to the fusion characteristics analogous to the nuclear structure studies. We have used the relativistic R3Y NN potential and densities from the RMF approach for the NL3$^*$ in the CCFULL calculation. The coupled channel calculations are performed with rotational degrees of freedom to examine the fusion enhancement using the microscopic nuclear potential at below-barrier energies. Firstly we perform the 1-dimensional barrier penetration model calculations by ignoring the nuclear intrinsic excitations to reproduce the experimental fusion cross-sections at energies above the barrier \cite{zamr06}. The black (solid), red (dashed) and blue (dotted) lines represent the fusion cross-sections obtained using WS, M3Y and R3Y NN interaction potential, respectively, for 1-D BPM as shown in Figs. \ref{fig2}, and \ref{fig3}. The calculated fusion cross-sections with 1-D BPM underestimate the experimental data, particularly at below-barrier energies. It is to be noted here that in the present CCFULL calculations, only rotational degrees of freedom are considered to address the fusion cross-section for the above-mentioned reactions. The two different conditions based on the shape (hexadecapole deformation) of nuclei $\beta_4 > 0 $, and $\beta_4 < 0$ are considered in the present work. Furthermore, it has been anticipated that the reactions involving nuclei either having positive or negative hexadecapole deformation exhibit an increase in the fusion cross-section at energies below the Coulomb barrier \cite{jain23,fern91,mort94,rhoa83,lemm93,fern89a}, which demands further experimental study.

The fusion cross-sections calculated for $^{16}$O + $^{24}$Mg, $^{18}$O + $^{24}$Mg, and $^{16}$O + $^{148}${Sm} reactions having $\beta_4 > 0$ are shown in Fig. \ref{fig2}. To probe the possible generalizations concerning the behaviour of the heavy-ion fusion process, firstly, we discuss the fusion reaction of $^{16,18}$O + $^{24}$Mg. Another aspect of nuclear structure that inspired the research of the $^{16,18}$O + $^{24}$Mg reaction is the significant deformation and strong collectively of $^{24}$Mg, which can potentially affect the fusion cross-section. For example, the size of the nuclear surface density distribution, where low-energy heavy-ion reactions occur, and the deformation might enhance the fusion cross-section. Furthermore, strong couplings to the collective degrees of freedom create a more efficient approach for kinetic energy transfer to internal excitation and influence the fusion cross-section. From Fig. \ref{fig2}(a), one can notice the fusion cross-section obtained for $^{16}$O + $^{24}$Mg reaction using WS potential slightly deviates from the M3Y interaction potential cross-section at near and below barrier energies. On the other hand, the fusion cross-section obtained with WS potential overlaps with the R3Y NN potential cross-section mainly at above barrier energies. Furthermore, the fusion cross-section obtained using R3Y NN interaction potential is comparatively larger than WS and M3Y interaction potential mainly at below-barrier energies.\\ \\
By including the excitation state of the target nuclei, the fusion cross-section is enhanced particularly at below barrier energies for M3Y (dotted-dashed green line) and R3Y (double-dotted dashed grey line) NN interaction potential and reasonably matches with the available experimental data \cite{tabo78}. This can be correlated with the weak magnitude of $\beta_4$ value causing the enhancement observed in the case of M3Y and R3Y NN interaction potential with the $(0^+ - 4^+)$ excitation state to be quite small compared to the 1-D BPM. Similar results can be pointed out for $^{18}$O + $^{24}$Mg reaction. The cross-section obtained using WS potential overlaps with the M3Y and R3Y NN interaction potential cross-section at energies above the Coulomb barrier. More detailed inspections show a small increase in the fusion cross-section for R3Y NN interaction potential in comparison to the WS and M3Y interaction potential and it is quite small for $(0^+ - 4^+)$ excitation state as compared to the 1-D BPM. Theoretical results obtained for these two reactions are consistent with the experimental data \cite{tabo78} over the available range of energies. More elaborately, the graph for the fusion cross-section is plotted for $^{16}$O + $^{24}$Mg and $^{18}$O + $^{24}$Mg reactions in the narrower energy range in the inset of Fig. \ref{fig2}(a), \ref{fig2}(b). The enhancement in the fusion cross-section with the inclusion of a rotational excitation state is more clearly visible at energies below the Coulomb barrier. As shown in Fig. \ref{fig2}(c) for $^{16}$O + $^{148}$Sm reaction, the data obtained using WS and M3Y interaction potential does not even fit the experimental cross-section \cite{leig95} data at the above barrier energies. However the calculated results with R3Y NN potential best match the experimental data near and above the Coulomb barrier energies. Also, the enhancement in the fusion cross-sections with the inclusion of $(0^+ - 4^+)$ excitation channel can be observed w.r.t. the 1-D BPM. It can be observed from the above-mentioned results with $\beta_4 > 0$ that the positive values of $\beta_4 > 0$ significantly affect the fusion cross-section and the results obtained using R3Y NN potential with rotational excitation state are slightly superior to the M3Y and WS potential for these considered systems. Further, with the inclusion of higher channels, a negligible effect on the fusion cross-sections is observed.\\ \\
A similar effect of the R3Y NN interaction nuclear potential can be observed for the reactions having negative values of the hexadecapole deformation of the target nuclei, $\beta_4 < 0$. The calculated fusion cross-section for all the considered system, namely, $^{16}$O + $^{176}${Hf}, $^{16}$O + $^{176}$Yb,  $^{16}$O + $^{182}${W}, $^{16}$O + $^{186}$W having $\beta_4 < 0$ with WS, M3Y and R3Y NN potential (without coupling terms) gives best match with the experimental data, especially at above barrier energies. However, fusion hindrance is still observed at energies below the Coulomb barrier. Therefore, the rotational excitation state of the target nuclei is included in the CC calculations. A detailed observation shows that the fusion cross-section obtained from M3Y potential exactly overlaps with the results obtained using WS potential at energies below and above the Coulomb barrier for $^{16}$O + $^{176}${Hf} as shown in Fig. \ref{fig3}(a). However, the fusion cross-section obtained using R3Y NN potential is more in comparison to the M3Y and WS potential, mainly at sub-barrier energies. The fusion cross-section obtained for the excitation state  $(0^+ - 4^+)$ is higher than the 1-D BPM for both M3Y and R3Y NN potential below barrier energies. While 1-D data overestimated the excitation fusion cross-section for M3Y potential at energies near and above the Coulomb barrier that related to the sharp cut-off model (choice of $\ell_{max}$-values). On the same footing, the results with R3Y NN potential match well with the experimental data \cite{leig88}  even at below-barrier energies. Moreover, similar calculations are done for other reactions namely $^{16}$O + $^{176}$Yb,  $^{16}$O + $^{182}${W}, $^{16}$O + $^{186}$W as shown in Fig. \ref{fig3}(b), \ref{fig3}(c), \ref{fig3}(d)\ref{fig3}(b), \ref{fig3}(c), \ref{fig3}(d). The experimental data are given for comparison \cite{rajb16,leig88,trot05}. From the figure, one can observe that for $\beta_4 < 0$, the significant effect of the $(0^+ - 4^+)$ excitation state on the fusion cross-section can be observed for R3Y NN potential as compared to the 1-D BPM with M3Y and WS potential. In other words, the fusion cross-sections obtained with the inclusion of the $(0^+ - 4^+)$ excitation state in R3Y NN potential are relatively closer to the experimental data at below and above barrier energies. As predicted above, we again observed the fusion cross-section obtained from the R3Y NN interaction potential with NL3$^*$ parameter set is found to be more consistent than that of WS and M3Y potential mainly at below-barrier energies, except $^{16}$O + $^{148}$Sm reaction at energies below the Coulomb barrier. In contrast, the M3Y interaction only fits the data at above-barrier energies. Notably, the R3Y nuclear interaction explains the fusion cross-section reasonably well at lower energies and consistently yields larger cross-sections than the M3Y potential. Furthermore, we observe that with the inclusion of higher-order channels beyond $4^+$, the higher-order channels cease to contribute significantly towards the fusion cross-section around and below the Coulomb barrier. The above analysis is evident from the Figs. \ref{fig2} and \ref{fig3} that the recently established R3Y interaction for NL3$^*$ force parameter with rotational degrees of freedom is a better choice than the WS and M3Y interaction potential for the considered systems in the fusion studies. Further, the implementation of nuclear potential from RMF in the coupled channel approach removes its dependency on the parameters of Wood-Saxon potential i.e. $V_{0}$, $r_{0}$ and $a_{0}$ as the potential here is obtained in a self-consistent way which makes the coupled channel model more reliable. The Woods-Saxon parameters are otherwise tuned manually at the above barrier energies for 1-D BPM.
\section{Summary and Conclusions}
\label{result} \noindent
In the present study, we have considered two different nucleon-nucleon interaction potentials- the widely used M3Y potential and the relatively new relativistic R3Y interaction to estimate the fusion characteristics at low energies. The nuclear interaction potential is obtained for relativistic R3Y and M3Y NN potential and corresponding interacting densities from the relativistic mean field approach for NL3$^*$ parameter set by adopting the double-folding procedure. The Coupled channels CCFULL code is used to calculate the fusion cross-section below barrier energies for these two kinds of nuclear potential and compared with the traditional Woods-Saxon (WS) potential. We considered seven reactions, namely, $^{16}$O + $^{24}$Mg,  $^{18}$O + $^{24}$Mg, $^{16}$O + $^{148}${Sm}, $^{16}$O + $^{176}${Hf}, $^{16}$O + $^{176}$Yb,  $^{16}$O + $^{182}${W}, $^{16}$O + $^{186}$W, in which target nuclei are rotational and $^{16}$O, $^{18}$O are considered as spherical in shape. In contrast to the expectation of 1-D BPM, the sub-barrier fusion cross-section is enhanced because of the coupling between the relative motion and the intrinsic degrees of freedom. The values of $\beta_2$ and $\beta_4$ deformation are calculated from relativistic mean-field formalism for NL3$^*$ parameter set used in the CCFULL to estimate the cross-section. Interestingly, even a small change in the barrier height of the R3Y potential has a significant impact on the cross-section, leading to a considerable increase in energies below the Coulomb barrier. From the fusion reactions at below barrier energies, we observed that the R3Y interaction for NL3$^*$ parameter set with rotational excitation state has proven to be a better option than the WS and M3Y potential. As a result, it can be concluded that the R3Y interaction with NL3$^*$ causes interacting nuclei to recline, which lowers the barrier and raises the cross-section at energies below the Coulomb barrier.

\section*{Acknowledgements}
This work has been supported by the Science Engineering Research Board (SERB) File No. CRG/2021/001229, Sao Paulo Research Foundation (FAPESP) Grant 2017/05660-0, and FOSTECT Project No. FOSTECT.2019B.04.


\end{document}